\documentclass[11pt,psfig]{article}
\usepackage{amsmath,amssymb,epsfig}

\newtheorem{theorem}{Theorem}[section]
\newtheorem{definition}[theorem]{Definition}

\newtheorem{lemma}[theorem]{Lemma}

\newtheorem{remark}[theorem]{Remark}

\newtheorem{assumption}[theorem]{Assumption}

\def\R{\mathbb{R}}

\def\P{\mathbb{P}}

\def\R{\mathbb{R}}

\def\P{\mathbb{P}}

\makeatletter 
\@addtoreset{equation}{section}

\makeatother 
\begin{document}
\begin{center}
{\Huge\bf On a Non-Standard Stochastic Control Problem}
\mbox{}\\
\vspace{2cm}
Ivar Ekeland
\footnote{Work supported by PIMS under NSERC grant 298427-04.}\\
Department of Mathematics\\
The University of British Columbia\\
Vancouver, BC, V6T1Z2\\
ekeland@math.ubc.ca\\

\vspace{1cm}
Traian A. Pirvu\\
Department of Mathematics\\
The University of British Columbia\\
Vancouver, BC, V6T1Z2\\
tpirvu@pims.math.ca\\
\vspace{1cm}

\mbox{}\\

\today
\end{center}

\noindent {\bf Abstract.}
 This paper considers the Merton portfolio management problem. We are concerned with non-exponential discounting of time
 and this leads to time inconsistencies of the decision maker. Following \cite{EkePir} we introduce the notion of equilibrium policies
 and we characterize them by an integral equation. The main idea is to come up with the \textit{value function} in this context. If risk preferences are of CRRA type, the integral equation which characterizes the \textit{value function} is shown to have a solution which leads to an equilibrium policy. This work is an extension of \cite{EkePir}.

\vspace{1cm}

\noindent {\bf Key words:} Portfolio optimization, Merton problem, Equilibrium policies, Non-exponential discounting

\begin{quote}

\end{quote}

\begin{flushleft}
{\bf JEL classification: }{G11}\\
{\bf Mathematics Subject Classification (2000): }
{91B30, 60H30, 60G44}
\end{flushleft}

\setcounter{equation}{0}
\section{Introduction}

This paper is a treatment of the Merton portfolio management problem in the context
 of non-exponential discounting. What makes this problem non-standard is that with non-exponential discounting, in the
absence of a commitment technology, the optimal policies computed as some time may fail
to remain optimal at later times. Therefore the notion of optimality is not robust across time in this paradigm. As it was
pointed out in \cite{EkePir} one way out of this predicament is to consider equilibrium policies. These are policies which
the decision maker at some point in time would have no incentive to infinitesimally deviate from them, given that at some later
times she agrees to follow them.

{\bf{Existing Research}}
In the financial-economic literature there is an interest in studying decision-making problems where the discount rate may not be
constant. Most of the results are in discrete time framework. The only paper in a continuous setup is \cite{EkeLaz}. For more on
this see \cite{EkePir}. 

Dynamic asset allocation in a Brownian motion model received allot of attention and the main paper in this area is due to Merton \cite{Mer71}.
In \cite{EkePir} we gave an overview of the works in the context of Merton portfolio management problem with exponential
discounting. In \cite{EkePir} we have a first attempt in addressing the issue of non-exponential discounting. By means of duality
we described the equilibrium policies by a flow of backward differential equations (BSDEs). In order to simplify the problem we
considered special cases of discounting for which it was possible to reduce the flow of BSDEs to systems of partial differential
equations PDEs. During this process we came up with a choice for the value function which was defined through an integral equation.
The duality method employed it turned out usefull, specially for the infinite horizon problem, in deriving transversality
like conditions. One of the novelty of \cite{EkePir} is the non-uniqueness of the equilibrium policies.

{\bf{Our Contributions}}

The present paper is a natural extension of \cite{EkePir}. We are able to use the value function approach
beyond the special cases of discounting considered in \cite{EkePir}. The key observation is that the
value function defined by an integral equation also satisfies an integral-differential equation which is
reminiscent of the classical Hamilton-Jacobi-Bellmann equation. However this equation has a non-local term due to
nonconstant discount rate. In the case of exponential discounting it coincides with the Hamilton-Jacobi-Bellmann equation. We
prove that a concave smooth solution of the integral equation characterizing the value function leads to an equilibrium policy. 
 This equation seems complicated and a general existence result is beyond the scope of
this work. However if one restricts to CRRA preferences existence of a smooth solution is established in this paper. In that case
the integral equation can be reduced to an one dimensional equation for which we proved existence and uniqueness. Moreover this
one dimensional equation is amenable to numerical treatments so one can compare the equilibrium policies arising from different
choices of discounting. We leave this as a topic of further research.

{\bf{Organization of the paper}}
The reminder of this paper is organized as follow.
 In section $2$ we describe the model and formulate the objective.
 Section $3$ contains the main result and is split in two parts: the model without the intertemporal
consumption and the model without the utility of the final wealth. Section $4$ treats the 
special case of CRRA type utilities. The paper ends with Appendix containing the proofs.

\section{The problem formulation}

\subsection{The model description}
 Following \cite{EkePir} we adopt a model for the financial market consisting of 
a saving account and one stock (risky asset). The inclusion of 
more risky assets can be achieved by notational changes. The saving
 account accrues interest at the riskless rate $r>0.$ The stock price
 per share follows an exponential Brownian motion
 $$
dS(t)=S(t)\left[\alpha\,dt
+\sigma\,dW(t)\right],\quad0\leq
t\leq\infty,
$$
where $\{W(t)\}_{t\in[0,\infty)}$ is a $1-$dimensional 
Brownian motion on a filtered probability space\\
$(\Omega,\{\mathcal{F}_t\}_{0\leq t\leq T},\mathcal{F},\P).$ 
The filtration $\{\mathcal{F}_t\}_{0\leq t\leq T}$ is the completed
 filtration generated by\\ $\{W(t)\}_{t\in[0,\infty)}.$
As usual $\alpha$ is \textit{ the mean rate of return}
and $\sigma>0$ is {\em the volatility}. Let us denote by
$\mu\triangleq\alpha-r>0$ {\em the excess return}.

\subsection{Investment-consumption policies and wealth processes}
A decision-maker in this market is continuously investing her 
wealth in the stock/bond and is consuming. An investment-consumption
 policy is determined by the proportion of current wealth he/she 
invests in the bond/stock and by the consumption rate. Formally we have:
\begin{definition}
\label{def:portfolio-proportions}
An $\R^{2}$-valued stochastic process
$\{\zeta(t), c(t)\}_{t\in[0,\infty)}$ is called an 
  admissible policy process if
it is progressively measurable, 
$c(t)\geq 0,\,\,\mbox{for all}\,\,t\in[0,\infty)$ and it
satisfies
\begin{equation}%
\label{kj**}
\int_0^t |\zeta(u)\mu-c(u)|\, du+\int_0^t
      |\zeta(u) \sigma|^2\, du<\infty, \text{ a.s., for all }
 t\in [0,\infty).
\end{equation}
\end{definition}

Given a policy process $\{\zeta(t), c(t)\}_{t\in[0,\infty)}$,
$\zeta(t)$ is the proportion of wealth, denoted by 
$X^{\zeta,c}(t),$ invested in the stock at time $t$
and $c(t)$ is the consumption rate. The
equation describing the dynamics of wealth ${X^{\zeta,c}(t)}$ is given by
\begin{eqnarray}\notag
dX^{\zeta,c}(t)&=&X^{\zeta,c}(t)\left((\alpha\zeta(t)-c(t))\,dt+
\sigma\zeta(t)\,dW(t)\right)
+(1-\zeta(t))X^{\zeta,c}(t)r\,dt\\\label{equ:wealth-one}&=&
X^{\zeta,c}(t)(\left(r+\mu\zeta(t)-c(t))\,dt
+\sigma\zeta(t)\,dW(t)\right).
\end{eqnarray}
It simply says that the changes in wealth over time are due solely
 to gains/loses from investing in stock, from consumption and there 
is no cashflow coming in or out. This is usually referred to
as the self-financing condition.

Under the regularity condition \eqref{kj} imposed on 
$\{\zeta(t), c(t)\}_{t\in[0,\infty)}$ above, equation
 \eqref{equ:wealth-one} admits a unique strong solution given by
the explicit expression
\begin{equation}\notag%
    \begin{split}
      X^{\zeta,c}(t)=X(0)\exp\left( \int_0^t
        \Big(r+\mu\zeta(u)-c(u)-\frac{1}{2}|\sigma\zeta(u)|^2\Big)\, du+
        \int_0^t \sigma\zeta(u)\, dW(u) \right),
    \end{split}
\end{equation}
The initial wealth $X^{\zeta,c}(0)=X(0)\in (0,\infty)$, 
is exogenously specified.

\subsection{Utility Function}

At all times the decision-maker has the same von Neumans-Morgenstern utility. 
This is crucial for understanding the model: time-inconsistency arise,
 not from a change in preferences, but from the way the future is 
discounted. The decision maker is deriving utility from intertemporal consumption and 
final wealth. Let $U$ be the utility of intertemporal consumption
 and $\hat{U}$ the utility of the terminal wealth
at some non-random horizon $T$ (which is a primitive of the model). Here the functions $U$ and $\hat{U},$ defined on $(0,\infty)\rightarrow \mathbb{R}$ are strictly increasing and strictly concave. We restrict ourselves to utility functions which are continuous differentiable and satisfy the Inada conditions
 \begin{equation}\label{In}
 U'(0+)\triangleq\lim_{x\downarrow 0}U'(x)=\infty,\quad U'(\infty)
\triangleq\lim_{x\uparrow{\infty}}U'(x)=0.
 \end{equation}
We shall denote by $I(\cdot)$ the (continuous, strictly decreasing) 
inverse of the marginal utility function $U'(\cdot),$
and by \eqref{In}
\begin{equation}\label{In1}
 I(0+)\triangleq\lim_{x\downarrow 0}I(x)=\infty,\quad I(\infty)
\triangleq\lim_{x\uparrow{\infty}}I(x)=0.
\end{equation}

\subsection{Discount Function}
Unlike other works in this area we do not restrict ourselves to the 
framework of exponential discounting. Following
\cite{EkeLaz}, a discount function $h:[0,\infty]\rightarrow
\mathbb{R}$  is assumed to be continuously differentiable, positive, with:
$$h(0)=1,\,\,h(s)\geq 0 ,$$
and
$$\int_{0}^{\infty}h(t)\,dt<\infty.$$
Moreover we assume that 
\begin{equation}\label{P&}
\inf_{t\in[0,T]}h(t)>0.
\end{equation}
 This implies that the discount rate function $-\frac{h'}{h}$ is bounded.

\subsection{The objective}
We conclude this section by formulating our problem. The objective
 is to find time consistent policies and the optimal ones may fail
 to have this feature. Indeed, as shown in \cite{EkePir} if the agent starts with a given 
positive wealth $x,$ at some instant $t,$ her optimal policy process
 $\{\tilde{\zeta}_{t}(s),\tilde{c}_{t}(s)\}_{s\in[t,T]}$ is chosen 
such that $$\sup_{\zeta,c}\mathbb{E}\left[\int_{t}^{T}h(u-t)U(c(u)
X^{\zeta,c}(u))\,du+h(T-t)\hat{U}(X^{\zeta,c}(T))\right]=$$
$$=\mathbb{E}\left[\int_{t}^{T}h(u-t)U(\tilde{c}(u)X^{\tilde{\zeta}
,\tilde{c}}(u))\,du+h(T-t)\hat{U}(X^{\tilde{\zeta},\tilde{c}}(T))\right].$$

The value function associated with this stochastic control problem is

\begin{equation}\notag
V(t,s,x)\triangleq\sup_{\zeta,c}\mathbb{E}\left[\int_{s}^{T}h(u-t)U(c(u)
X^{\zeta,c}(u))\,du+h(T-s)\hat{U}(X^{\zeta,c}(T))\bigg|X(s)=x\right],
\end{equation}

$t\leq s\leq T,$ and it solves the following Hamilton-Jacobi-Bellman equation
$$
\frac{\partial V}{\partial s}(t,s,x)+\sup_{\zeta,c}\left[(r+\mu\zeta-c)x
\frac{\partial V}{\partial x}(t,s,x)+\frac{1}{2}\sigma^{2}\zeta^{2}x^{2}
\frac{\partial^{2} V}{\partial x^{2}}(t,s,x)\right]$$$$+
\frac{h'(s-t)}{h(s-t)}V(t,s,x)+U(xc)=0,
$$
with the boundary condition
\begin{equation}\label{boundarycondition}
V(t,T,x)=\hat{U}(x).
\end{equation}

The first order necessary conditions yield the $t-$optimal policy $\{\tilde{\zeta}
_{t}(s),\tilde{c}_{t}(s)\}_{s\in[t,T]}$
\begin{equation}\label{11}
\tilde{\zeta}_{t}(s,x)=-\frac{ \mu\frac{\partial V}{\partial x}(t,s,x)
}{\sigma^{2}x\frac{\partial^{2} V}{\partial x^{2}}(t,s,x)},\quad t\leq s\leq T,
\end{equation}
\begin{equation}\label{c11}
\tilde{c}_{t}(s,x)=\frac{I(\frac{\partial V}{\partial x}(t,s,x))}{x},
\quad t\leq s\leq T.
\end{equation}

Therefore, unless the discounting is exponential (in which case 
$\frac{h'}{h}=\mbox{constant},$ so there is no $t$
dependence in the HJB), the $t-$optimal policy may not be optimal
 after $t.$ That is $$ \{\tilde{\zeta}_{t}(s),\tilde{c}_{t}(s)\}
_{s\in[t',T]}\neq\arg\max_{\zeta, c}\mathbb{E}\left[\int_{t'}^{T}
h(u-t')U(c(u)X^{\zeta,c}(u))\,du+h(T-t')\hat{U}(X^{\zeta,c}(T))\right],$$
for some subsequent instant $t',$ so the decision-maker would implement 
the $t-$optimal policy  at later times only if she is constrained to 
do so.

The resolution is to introduce equilibrium policies,
that is, policies which, given that they will be implemented in the future, 
it is optimal to implement them right now.  Let us consider:

\begin{equation}\label{eq}
\bar{\zeta}(s,x)=\frac{F_{1}(s,x)}{x},\quad\bar{c}(s,x)=\frac{F_{2}(s,x)}{x},
\end{equation}
for some functions $F_{1},$ $F_{2}.$ The corresponding wealth process 
$\{\bar{X}(s)\}_{s\in[0,T]}$ evolves according to
\begin{equation}\label{dyn}
d\bar{X}(s)=[r\bar{X}(s)+\mu F_{1}(s,\bar{X}(s))-F_{2}(s,\bar{X}(s))]ds+
\sigma F_{1}(s,\bar{X}(s))dW(s).
\end{equation} 
Following \cite{EkePir}  the functions $F_{1}, F_{2}$ will be chosen such
 that on $[t,t+\epsilon]$ it is optimal (this is made precise
in our formal definition of equilibrium policies) to pick
$\bar{\zeta}(t,x)=\frac{F_{1}(t,x)}{x},\quad\bar{c}(t,x)=\frac{F_{2}(t,x)}{x},$
 given that the agent's wealth at time $t$ is $x,$
and that for every subsequent instance $s\geq t+\epsilon$ she follows \eqref{eq}.
This intuition will be made precise in the following sections.

\section{The Main Result}

\subsection{The equilibrium policies and the value function}
Let $T$ be a finite time horizon exogenously specified.
In general, for a policy process\\ $\{{\zeta}(s),{c}(s)\}_{s\in[0,T]}$
 satisfying \eqref{kj**} and its corresponding wealth process 
$\{X(s)\}_{s\in[0,T]}$\\ (see \eqref{equ:wealth-one})
we denote the expected utility functional by
\begin{equation}\label{01FUNCT}
J(t,x,\zeta,c)\triangleq\mathbb{E}\left[\int_{t}^{T}h(s-t)U(c(s)X^{t,x}(s))
\,ds+h(T-t)\hat{U}(X^{t,x}(T))\right].
\end{equation}
 Following \cite{EkePir} we shall give a rigorous mathematical
formulation of the equilibrium policies in the 
formal definition below.

\begin{definition}\label{finiteh}
A map $F=(F_{1},F_{2}):(0,\infty)\times[0,T]\rightarrow\mathbb{R}
\times[0,\infty)$ is an equilibrium policy
for the finite horizon investment-consumption problem, if for any $t,x>0$
\begin{equation}\label{opt}
{\lim\inf_{\epsilon\downarrow 0}}\frac{J(t,x,F_{1},F_{2})-
J(t,x,\zeta_{\epsilon},c_{\epsilon})}{\epsilon}\geq 0,
\end{equation}
where
$$J(t,x,F_{1},F_{2})\triangleq J(t,x,\bar{\zeta},\bar{c}),$$
\begin{equation}\label{0000eq}
\bar{\zeta}(s)=\frac{F_{1}(s,\bar{X}(s))}{\bar{X}(s)},
\quad\bar{c}(s)=\frac{F_{2}(s,\bar{X}(s))}{\bar{X}(s)},
\end{equation}
and $\{\bar{\zeta}(s),\bar{c}(s)\}_{s\in[t,T]}$  satisfies \eqref{kj}.
 The equilibrium wealth process $\{\bar{X}(s)\}_{s\in[t,T]}$ is a 
solution of the stochastic differential equation (SDE)
\begin{equation}\label{0dyn}
d{X}(s)=[r{X}(s)+\mu F_{1}(s,{X}(s))-F_{2}(s,{X}(s))]ds+\sigma F_{1}(s,{X}(s))dW(s).
\end{equation}

The process $\{{\zeta}_{\epsilon}(s),{c}_{\epsilon}(s)\}_{s\in[t,T]}$
 is another investment-consumption policy defined by
\begin{equation}\label{1e}
\zeta_{\epsilon}(s)=\begin{cases} \bar{\zeta}(s),\quad s\in[t,T]\backslash E_{\epsilon,t}\\
\zeta(s), \quad s\in E_{\epsilon,t}, \end{cases}
\end{equation}

\begin{equation}\label{2e}
c_{\epsilon}(s)=\begin{cases} \bar{c}(s),\quad s\in[t,T]\backslash E_{\epsilon,t}\\
c(s), \quad s\in E_{\epsilon,t}, \end{cases}
\end{equation}
with $E_{\epsilon,t}=[t,t+\epsilon],$ and  $\{{\zeta}(s),{c}(s)\}_{s\in 
E_{\epsilon,t} }$ is any policy for which $\{{\zeta}_{\epsilon}(s),{c}_{\epsilon}
(s)\}_{s\in[t,T]}$ is an admissible policy.
\end{definition}

In \cite{EkePir} in a first step we characterize the equilibrium policies by means of duality.
This methodology was illuminating and inspired us for a value function candidate. Thus following
\cite{EkePir}, the value function $v$ should solve the following equation
\begin{equation}\label{00ie1}
v(t,x)=\mathbb{E}\left[\int_{t}^{T}h(s-t)U(F_{2}(s,\bar{X}^{t,x}(s)))\,ds+
h(T-t)\hat{U}(\bar{X}^{t,x}(T))\right],
\end{equation}
 Recall that $\{\bar{X}(s)\}_{s\in[0,T]}$
is the equilibrium wealth process and it satisfies
\begin{equation}\label{10dyn}
d\bar{X}(s)=[r\bar{X}(s)+\mu F_{1}(s,\bar{X}(s))-F_{2}(s,\bar{X}(s))]ds+
\sigma F_{1}(s,\bar{X}(s))dW(s),
\end{equation}
where $F=(F_{1},F_{2})$ is given by
 \begin{equation}\label{109con}
 F_{1}(t,x)=-\frac{\mu\frac{\partial v}{\partial x}(t,x)}{\sigma^{2}
\frac{\partial^{2} v}{\partial x^{2}}(t,x)},\,\,
F_{2}(t,x)=I\left(\frac{\partial v}{\partial x}(t,x)\right),\,\,\,t\in[0,T].
\end{equation}

Of course the existence of such a function $v$ satisfying the equation above is not
a trivial issue. We can address it in some special cases (CRRA utilities for example). It turns out
that the map $F=(F_{1},F_{2})$ is an equilibrium policy. The next Theorem which is the central result
of the paper asserts this.

\begin{theorem}\label{main}
Assume there exists a concave function $v$ of class $C^{1,2}$ which satisfies \eqref{00ie1} together with \eqref{10dyn} and \eqref{109con}.
Under the Assumptions \ref{As}, or \ref{As1} and \ref{As2}, the map $F=(F_{1},F_{2})$ given by \eqref{109con} is an equilibrium policy.
\end{theorem}

Appendix A proves this Theorem. 

\begin{flushright}
$\square$
\end{flushright}

In the following, to ease to exposition, we present the two polar cases:
\begin{itemize}
 \item The case without intertemporal consumption: $U=0.$  
 
 \item The case without the utility of the final wealth: $\hat{U}=0.$
\end{itemize}

\subsection{The case without intertemporal consumption}
In this case $F_{2}=0$ and we denote $F_{1}\triangleq F.$ This case is easier to handle because
the ``aggregate'' term is missing. The integral equation now becomes
\begin{equation}\label{000ie1}
v(t,x)=\mathbb{E} [h(T-t)\hat{U}(\bar{X}^{t,x}(T))],
\end{equation}
where $\{\bar{X}(s)\}_{s\in[0,T]}$
is the equilibrium wealth process defined by the SDE
\begin{equation}\label{710dyn}
d{X}(s)=[r{X}(s)+\mu F(s,{X}(s))]ds+\sigma F(s,{X}(s))dW(s),
\end{equation}
and

\begin{equation}\label{0109con}
 F(t,x)=-\frac{\mu\frac{\partial v}{\partial x}(t,x)}{\sigma^{2}
\frac{\partial^{2} v}{\partial x^{2}}(t,x)}.
\end{equation}

The next Lemma shows that a PDE equation is the analogue of the integral equation \eqref{000ie1}.

\begin{lemma}\label{pDe}
 If a function $v$ of class $C^{1,2}$ satisfies the integral equation \eqref{000ie1}, then $v$ solves the following PDE
\begin{equation}\label{a1}
\frac{\partial v}{\partial t}(t,x)+\frac{h'(T-t)}{h(T-t)}v(t,x)+rx\frac{\partial v}{\partial x}(t,x)-
\frac{\mu^{2}}{2\sigma^{2}}\frac{{[\frac{\partial v}{\partial x}}(t,x)]^{2}}{\frac{\partial^{2} v}{\partial x^{2}}(t,x)}
=0,
\end{equation}
with the boundary condition

\begin{equation}\label{0boundarycondition}
v(T,x)=\hat{U}(x).
\end{equation}

\end{lemma}

Appendix B proves this Theorem.
\begin{flushright}
$\square$
\end{flushright}

The existence of a solution for the PDE equation \eqref{a1} is established in the next Lemma. 

\begin{lemma}\label{0pDe}
There exists a unique concave $C^{1,2}$ solution (with exponential growth) of the PDE \eqref{a1} with the
boundary condition \eqref{0boundarycondition}.
\end{lemma}

Appendix C proves this Theorem.

\begin{flushright}
$\square$
\end{flushright}

The solution of the PDE \eqref{a1}, with the
boundary condition \eqref{0boundarycondition} leads to equilibrium policies. However some technical assumptions
are needed.

\begin{assumption}\label{As}
Given $v$ the $C^{1,2}$ solution of the PDE \eqref{a1}, with the
boundary condition  \eqref{0boundarycondition} and the map $F$ of \eqref{0109con}, 
we assume that the SDE \eqref{710dyn} has a strong solution. Moreover assume that the process $\int_{0}^{t}\frac{\partial v}{\partial x}(s,\bar{X}(s))
\sigma F(s,\bar{X}(s))\,dW(s)$ is a (true) martingale.
\end{assumption}

In the following we restrict ourselves to investment-consumption policies $\{\zeta(t), c(t)\}_{t\in[0,\infty)}$ as in the Definition \ref{def:portfolio-proportions}
such that the process
\begin{equation}\label{*89}
  \int_{0}^{t}\frac{\partial v}{\partial x}(s,{X}^{\zeta,c}(s))
\sigma\zeta(s)\,dW(s),\quad\mbox{is\,\,a\,\, martingale}.
\end{equation}

\begin{theorem}\label{main1}
Assume that $v$ the $C^{1,2}$ solution of the PDE \eqref{a1}, with the boundary condition \eqref{0boundarycondition} satisfies Assumption \ref{As}. Then the map $F$ of \eqref{0109con} is an 
equilibrium policy.
\end{theorem}

Appendix D proves this Theorem.

\begin{flushright}
$\square$
\end{flushright}

\begin{remark}
 The technical Assumption \eqref{As} is met for power utilities $U_{p}(x)=\frac{x^{p}}{p}.$ In this case the value function $v$ is of the form 
\begin{equation}\label{oP}
 v(t,x)=\lambda(t)U_{p}(x),
\end{equation}
 for a function $\lambda$ which solves the ODE
\begin{equation}\label{oDe}
\lambda'(t)+\left[\frac{h'(T-t)}{h(T-t)}+rp+\frac{\mu^2 p}{2\sigma^2 (1-p)} \right]\lambda(t)=0,\qquad \lambda(T)=1.
\end{equation}
According to Lemma \eqref{0pDe} we know that $v$ is the unique solution of the PDE \eqref{a1}, with boundary condition \eqref{0boundarycondition}, so it must be of the form
\eqref{oP}. Therefore the equilibrium policy is
\begin{equation}\label{eQ}
 F(t,x)=\frac{\mu x}{\sigma^2 (1-p)},
\end{equation}
hence the Assumption \eqref{As} hold true. Indeed it suffices to prove that
$$\mathbb{E}\int_{0}^{T}|\bar{X}(s)|^{2p}\,ds<\infty.$$
This follows from $\mathbb{E}|\bar{X}(s)|^{2p}=e^{K' s},$ where $K'=p\left(r+\frac{\mu^{2}}{2(1-p)^{2}\sigma^{2}}\right).$
As for the admissible policies uniformly bounded $(\zeta,c)$ will satisfy the restriction \eqref{*89}.

\end{remark}

\begin{remark}
 Notice that for the power utilities the equilibrium policy is to invest a constant proportion
(which depends on the discount rate) into the risky asset. This is a special case and it has to do
partly with stock return and volatilities being constant on one hand and on the other hand with the
particular choice of utility. Notice that the optimal policies do not depend on the discount rate,
since this being deterministic can be factored outside of the expectation operator.
\end{remark}

\subsection{The case without the utility of final wealth}

In this case set $\hat{U}=0.$ The value function
 $v(t,x)$ should satisfy the integral equation
\begin{equation}\label{0ie1}
v(t,x)=\mathbb{E}\left[\int_{t}^{T}h(s-t)U(F_{2}(s,\bar{X}^{t,x}(s)))\,ds \right].
\end{equation}
 Recall that $\{\bar{X}(s)\}_{s\in[0,T]}$
is the equilibrium wealth process and it follows the SDE
\begin{equation}\label{*010dyn}
d\bar{X}(s)=[r\bar{X}(s)+\mu F_{1}(s,\bar{X}(s))-F_{2}(s,\bar{X}(s))]ds+\sigma F_{1}(s,\bar{X}(s))dW(s),
\end{equation}

with

\begin{equation}\label{*109con}
 F_{1}(t,x)=-\frac{\mu\frac{\partial v}{\partial x}(t,x)}{\sigma^{2}\frac{\partial^{2} v}{\partial x^{2}}(t,x)},\,\,
F_{2}(t,x)=I\left(\frac{\partial v}{\partial x}(t,x)\right),\,\,\,t\in[0,T].
\end{equation}

The following Lemma derives a integro-differential equation analogue to the integral equation \eqref{0ie1}. We need the following
technical assumption.

\begin{assumption}\label{As1}
For every $s\in[0,T]$ the PDE, defined on $[0,s]\times(0,\infty),$
\begin{equation}\label{PDE1*}
f_{t}+(r x+\mu F_{1}(t,x)-F_{2}(t,x))f_{x}+\frac{\sigma^{2} F_{1}^{2}(t,x)}{2} f_{xx}=0,\quad f(s,s,x)=U(F_{2}(s,x)),
\end{equation}
has a $C^{1,2}$ solution $f=f(t,s,x).$
\end{assumption}

\begin{lemma}\label{De}
Assume there exists a $C^{1,2}$ solution $v=v(t,x)$ of the integral equation \eqref{0ie1} and that the Assumption \ref{As1} is satisfied. Then $v(t,x)$ solves the following integro-differential equation
\begin{equation}\label{1dE}
 \frac{\partial v}{\partial t}(t,x)+\left(rx-I\left(\frac{\partial v}{\partial x}(t,x)\right) \right)\frac{\partial v}{\partial x}(t,x)-
\frac{\mu^{2}}{2\sigma^{2}}\frac{{[\frac{\partial v}{\partial x}}(t,x)]^{2}}{\frac{\partial^{2} v}{\partial x^{2}}(t,x)}+U(F_{2}(t,x))=
\end{equation}

$$-\mathbb{E}\left[\int_{t}^{T}h'(s-t)U\left(I\left(\frac{\partial v}{\partial x}(s,\bar{X}^{t,x}(s))\right)\right)\,ds\right].$$  
\end{lemma}

Appendix E proves this Lemma.

\begin{flushright}
$\square$
\end{flushright}

In the following we restrict ourselves to investment-consumption policies $\{\zeta(t), c(t)\}_{t\in[0,\infty)}$ as in the Definition \ref{def:portfolio-proportions}
such that the process
\begin{equation}\label{89}
  \int_{0}^{t}\frac{\partial v}{\partial x}(s,{X}^{\zeta,c}(s))
\sigma\zeta(s)\,dW(s),\quad\mbox{is\,\,a\,\, martingale}.
\end{equation}
Moreover we require that
\begin{equation}\label{189}
 \mathbb{E} \sup_{\{0\leq s\leq T\}} |U(c(s){X}^{\zeta,c}(s))|<\infty,
\end{equation}

and

\begin{equation}\label{*289}
 {\lim_{\epsilon\downarrow 0}}\frac{\mathbb{E}\int_{t}^{t+\epsilon}\bigg[\frac{\partial v}{\partial t}(s,{X}^{\zeta,c}(s)))+\bigg[(r+\mu(s)\zeta(s)-c(s)){X}^{\zeta,c}(s))\frac{\partial v}{\partial x}(s,{X}^{\zeta,c}(s)))\bigg]+}{\epsilon}
\end{equation}

\begin{equation*}\label{289}
\frac{+\bigg[\frac{1}{2}\sigma^{2}(s)\zeta^{2}(s)({X}^{\zeta,c}(s))^{2}\frac{\partial^{2} v}{\partial x^{2}}(s,{X}^{\zeta,c}(s)))\bigg]\bigg]ds}{\epsilon}
\end{equation*}

\begin{equation*}
= \frac{\partial v}{\partial t}(t,x)+\left[(r+\mu(t)\zeta(t)-c(t))x\frac{\partial v}{\partial x}(t,x)+\frac{1}{2}\sigma^{2}\zeta^{2}x^{2}\frac{\partial^{2} v}{\partial x^{2}}(t,x)\right]
\end{equation*}

 At this point another technical assumption is imposed.

\begin{assumption}\label{As2}
Assume there exists a $C^{1,2}$ solution $v=v(t,x)$ of the integral equation \eqref{0ie1}, that the SDE \eqref{*010dyn} has a strong solution, and that the process $\int_{0}^{t}\frac{\partial v}{\partial x}(s,\bar{X}(s))
\sigma F(s,\bar{X}(s))\,dW(s)$ is a (true) martingale. Moreover assume that

\begin{equation}\label{089}
 \mathbb{E} \sup_{\{0\leq s\leq T\}}|U(F_{2}(s,\bar{X}(s)))|<\infty,
\end{equation}
and that the policies 
\begin{equation}\label{0eq}
\bar{\zeta}(s,x)=\frac{F_{1}(s,x)}{x},\quad\bar{c}(s,x)=\frac{F_{2}(s,x)}{x},
\end{equation}
satisfies \eqref{*289}.

\end{assumption}

\begin{theorem}\label{main2}
Assume there exists a concave function $v$ of class $C^{1,2}$ which satisfies the integral equation \eqref{0ie1}
and the Assumptions \ref{As1}, \ref{As2} hold true. Then the map $F=(F_{1},F_{2})$ given by \eqref{*109con} is an equilibrium policy.
\end{theorem}

Appendix F proves this Theorem.

\begin{flushright}
$\square$
\end{flushright}

\section{CRRA Preferences}

In the case of $U(x)=\hat{U}(x)=U_{p}(x)=\frac{x^{p}}{p},$ let us look for the value function $v$ of the form $ v(t,x)=\lambda(t)U_{p}(x),$
 for a function $\lambda(t)$ which is to be found. Notice that
$$F_1(t,x)=\frac{\mu x}{\sigma^{2}(1-p)},\,\,\,\,\,\,\,\,\,\,\,\,F_{2}(t,x)=[\lambda(t)]^{\frac{1}{p-1}}x.$$

 In the light of \eqref{0ie1} the function
$\lambda(t)$ should satisfy the following integral equation (IE)
\begin{equation}\label{IE0}
\begin{cases}
\lambda(t)=\int_{t}^{T} h(s-t)e^{K(s-t)} [\lambda(s)]^{\frac{p}{p-1}}e^{-\left(\int_{t}^{s}p[\lambda(u)]^{\frac{1}{p-1}}\,du\right)}\,ds+
h(T-t)e^{K(T-t)}e^{-\left(\int_{t}^{T}p[\lambda(u)]^{\frac{1}{p-1}}\,du\right)}\\\lambda(T)=1\end{cases}
\end{equation}
where $K=p\left(r+\frac{\mu^{2}}{2(1-p)\sigma^{2}}\right).$ The following Theorem grants existence and uniqueness for the solution of the above equation.

\begin{theorem}\label{ain2}
 There exists a unique $C^{1}$ solution of \eqref{IE0}.
\end{theorem}

Appendix G proves this Theorem.

\begin{flushright}
$\square$
\end{flushright}

\begin{remark}
In the context of CRRA preferences one can easily see that the Assumptions \ref{As1} and \ref{As2} hold true.
  Indeed in this case the PDE \eqref{PDE1*} is just the classical Black-Scholes PDE hence assumption \ref{As1} is met. As for
the Assumption \ref{As2} notice that the equilibrium wealth process is Gaussian. The class of uniformly bounded
policies $(\zeta,c)$ are admissible in the sense that the restrictions \eqref{89}, \eqref{189} and \eqref{*289} are met
for this class. 
\end{remark}

\section{Appendix}

{\bf A Proof of Theorem \ref{main}} The proof is a combination of Theorem \ref{main1} and Theorem \ref{main2}'s proofs.

\begin{flushright}
$\square$
\end{flushright}

{\bf B Proof of Lemma \ref{pDe}}

Proof: If a function $v$ satisfies the integral equation \eqref{000ie1} then the process $\frac{v(t,\bar{X}(t))}{h(T-t)}$ is a martingale
and $v(T,x)=\hat{U}(x)$. The result follows from It\^{o}'s formula.

\begin{flushright}
$\square$
\end{flushright}

{\bf C Proof of Lemma \ref{0pDe}}

Proof: Let us introduce the 
Legendre-Fenchel transform of $\hat{U}(x)$
\begin{equation}\label{Leg}
\hat{U}^{*}(y)\triangleq\sup_{x>0}[\hat{U}(x)-xy],\quad 0<y<\infty.
\end{equation}
The function $\hat{U}^{*}(\cdot)$ is strictly decreasing, strictly convex and satisfies 
 the dual relationships
\begin{equation}\label{In12}
\hat{U}^{*'}(y)=-x\quad\mbox{iff}\quad \hat{U}'(x)=y.
 \end{equation}
Let $\tilde{v}$ the unique (with exponential growth) convex $C^{1,2}$ solution of
\begin{equation}\label{a2}
\frac{\partial \tilde{v}}{\partial t}(t,y)+\frac{h'(T-t)}{h(T-t)}[y\frac{\partial\tilde{v}}{\partial y}(y,t)+\tilde{v}(t,y)]+ry\frac{\partial\tilde{v}}{\partial y}(y,t)+
\frac{\mu^{2}}{2\sigma^{2}}y^{2} \frac{\partial^{2} \tilde{v}}{\partial y^{2}}(t,y)
=0,
\end{equation}
with the boundary condition

\begin{equation}\label{00boundarycondition}
\tilde{v}(T,y)=\hat{U}^{*}(y).
\end{equation}
Note that the convexity of the function $\tilde{v}$ is inherited from its boundary condition. Let
\begin{equation}\label{kj**1}
{v}(t,x)=\inf_{y>0}[xy+\tilde{v}(t,y)].
\end{equation}
The following \textit{dual} relations hold
\begin{equation}\label{lp}
\frac{\partial\tilde{v}}{\partial y}(t,y)=-x,\quad\mbox{iff}\quad \frac{\partial {v}}{\partial x}(t,x)=y, 
\end{equation}
and
\begin{equation}\label{i9}
\frac{\partial^{2}\tilde{v}}{\partial y^{2}}(t,y)=-\frac{1}{\frac{\partial^{2} {v}}{\partial x^{2}}(t,x)},
\end{equation}
where $x$ and $y$ are related by \eqref{lp}. Moreover by the Envelope Theorem $\frac{\partial\tilde{v}}{\partial t}(y,t)
=\frac{\partial {v}}{\partial x}(x,t).$ Then it is easy to see that $v$ defined by \eqref{kj**1} solves \eqref{a1}. Moreover the concavity of $v$ follows from \eqref{i9}. Conversely if $v$ is a $C^{1,2}$ solution of \eqref{a1} then the function
\begin{equation}\label{kj}
\tilde{v}(t,x)=\sup_{x>0}[{v}(t,x)-xy].
\end{equation}
solves \eqref{a2}.

\begin{flushright}
$\square$
\end{flushright}

{\bf D Proof of Theorem \ref{main1}}

The claim yields if we
prove that
$$\mathbb{E} [h(T-t)\hat{U}(\bar{X}^{t,x}(T))]\geq \mathbb{E} [h(T-t)\hat{U}({X}^{\epsilon,t,x}(T))]. $$
This follows since the process $\left\{\frac{v(s,\bar{X}(s))}{h(T-s)}\right\}_{s\in[t,T]}$ is a martingale,
and the process $\left\{\frac{v(s,{X}^{\epsilon}(s))}{h(T-s)}\right\}_{s\in[t,T]}$ is a submartingale.
Indeed it can be seen by applying It\^{o}'s formula and using concavity of $v,$ that $\left\{\frac{v(s,\bar{X}(s))}{h(T-s)}\right\}_{s\in[t,T]}$ has zero drift and $\left\{\frac{v(s,{X}^{\epsilon}(s))}{h(T-s)}\right\}_{s\in[t,T]}$ has negative drift.

\begin{flushright}
$\square$
\end{flushright}

{\bf E Proof of Lemma \ref{De}}

 Equation \eqref{0ie1}
can be rewritten as
\begin{equation}\label{1ie1}
v(t,x)=\int_{t}^{T}h(s-t) f(t,s,x)\,ds,
\end{equation}

where

$$f(t,s,x)\triangleq\mathbb{E}[U(F_{2}(s,\bar{X}^{t,x}(s)))].$$

For a fixed $s$ the process $\{f(t,s,\bar{X}(t))\}_{0\leq t\leq s}$
is a martingale. Thus by the Assumption \ref{As1} the function $f=f(t,s,x)$ satisfies the following PDE:
\begin{equation}\label{PDE1}
f_{t}+(r x+\mu F_{1}(t,x)-F_{2}(t,x))f_{x}+\frac{\sigma^{2} F_{1}^{2}(t,x)}{2} f_{xx}=0.
\end{equation}

By differentiating  \eqref{1ie1} with respect to $t$ we get

\begin{equation}\label{t1ie1}
v_{t}(t,x)=\int_{t}^{T}h(s-t) f_{t}(t,s,x)\,ds-\int_{t}^{T}h'(s-t) f(t,s,x)\,ds.
\end{equation}

Moreover

\begin{equation}\label{xie1}
v_{x}(t,x)=\int_{t}^{T}h(s-t) f_{x}(t,s,x)\,ds.
\end{equation}
 
\begin{equation}\label{yie1}
v_{xx}(t,x)=\int_{t}^{T}h(s-t) f_{xx}(t,s,x)\,ds.
\end{equation}

In the light of \eqref{PDE1}, \eqref{t1ie1}, \eqref{xie1} and \eqref{yie1} it follows that

$$v_{t}(t,x)+(r x+\mu F_{1}(t,x)-F_{2}(t,x))v_{x}(t,x)+\frac{\sigma^{2} F_{1}^{2}(t,x)}{2} v_{xx}(t,x)+U(F_{2}(t,x))= $$
$$=-\int_{t}^{T}h'(s-t) f(t,s,x)\,ds,$$

or equivalently

\begin{equation}\label{01dE}
 \frac{\partial v}{\partial t}(t,x)+\left(rx-I\left(\frac{\partial v}{\partial x}(t,x)\right) \right)\frac{\partial v}{\partial x}(t,x)-
\frac{\mu^{2}}{2\sigma^{2}}\frac{{[\frac{\partial v}{\partial x}}(t,x)]^{2}}{\frac{\partial^{2} v}{\partial x^{2}}(t,x)}+U(F_{2}(t,x))=
\end{equation}

$$-\mathbb{E}\left[\int_{t}^{T}h'(s-t)U\left(I\left(\frac{\partial v}{\partial x}(s,\bar{X}^{t,x}(s))\right)\right)\,ds\right].$$

\begin{flushright}
$\square$
\end{flushright}

{\bf F Proof of Theorem \ref{main2}}

According to Lemma \ref{De}, $v$ satisfies the integro-differential equation

\begin{equation}\label{2dE}
 \frac{\partial v}{\partial t}(t,x)+\left(rx-I\left(\frac{\partial v}{\partial x}(t,x)\right) \right)\frac{\partial v}{\partial x}(t,x)-
\frac{\mu^{2}}{2\sigma^{2}}\frac{{[\frac{\partial v}{\partial x}}(t,x)]^{2}}{\frac{\partial^{2} v}{\partial x^{2}}(t,x)}+U(F_{2}(t,x))=
\end{equation}

$$-\mathbb{E}\left[\int_{t}^{T}h'(s-t)U\left(I\left(\frac{\partial v}{\partial x}(s,\bar{X}^{t,x}(s))\right)\right)\,ds\right].$$

In view of the concavity of function $v$ this equation can be re-written as
\begin{eqnarray}\label{io}
\frac{\partial v}{\partial t}(t,x)+\sup_{\zeta,c}\left[(r+\mu\zeta-c)x\frac{\partial v}{\partial x}(t,x)+\frac{1}{2}\sigma^{2}\zeta^{2}x^{2}\frac{\partial^{2} v}{\partial x^{2}}(t,x)+U(xc)\right]\\\notag=
-\mathbb{E}\left[\int_{t}^{T}h'(s-t)U(F_2(s,\bar{X}^{t,x}(s)))\,ds\right].
\end{eqnarray}

Let us notice that

$$J(t,x,F_{1},F_{2})=v(t,x).$$

Moreover

$$J(t,x,F_{1},F_{2})-J(t,x,\zeta_{\epsilon},c_{\epsilon})=\mathbb{E}\left[\int_{t}^{t+\epsilon}h(s-t)[U(F_2(s,\bar{X}^{t,x}(s)))-U(c(s)\bar{X}^{t,x}(s))]\,ds\right]$$
$$+\mathbb{E}[v(t+\epsilon,\bar{X}^{t,x}(t+\epsilon))-v(t+\epsilon,{X}^{t,x}(t+\epsilon))]$$
$$+\mathbb{E}\left[\int_{t+\epsilon}^{T}[h(s-t)-h(s-t-\epsilon)][U(F_2(s,\bar{X}^{t,x}(s)))-U(F_2(s,{X}^{t,x}(s)))]\,ds\right]. $$

The RHS of this equation has three terms and we will treat each of these terms separately:

$1.$ In the light of inequalities \eqref{089} and \eqref{189} and Dominated Convergence Theorem

$${\lim_{\epsilon\downarrow 0}}\frac{\mathbb{E}\left[\int_{t}^{t+\epsilon}h(s-t)[U(F_2(s,\bar{X}^{t,x}(s)))-U(c(s)X^{t,x}(s))]\,ds\right]}{\epsilon}=
U(F_2(t,x)-U(cx).$$

$2.$ One has

$$\mathbb{E}[v(t+\epsilon,\bar{X}^{t,x}(t+\epsilon))-v(t+\epsilon,{X}^{t,x}(t+\epsilon))]=$$
$$\mathbb{E}[v(t+\epsilon,\bar{X}^{t,x}(t+\epsilon))-v(t,x)]-\mathbb{E}[v(t+\epsilon,{X}^{t,x}(t+\epsilon)-v(t,x)].$$
Moreover
$$\mathbb{E}[v(t+\epsilon,\bar{X}^{t,x}(t+\epsilon))-v(t,x)]=\mathbb{E}\int_{t}^{t+\epsilon}d v(u, \bar{X}^{t,x}(u)).$$
It\^{o}'s formula and Assumption \ref{As2} yield

$${\lim_{\epsilon\downarrow 0}}\frac{\mathbb{E}\int_{t}^{t+\epsilon}d v(u, \bar{X}^{t,x}(u))}{\epsilon}= 
\frac{\partial v}{\partial t}(t,x)+\left(rx-I\left(\frac{\partial v}{\partial x}(t,x)\right)\right)\frac{\partial v}{\partial x}(t,x)-
\frac{\mu^{2}}{2\sigma^{2}}\frac{{[\frac{\partial v}{\partial x}}(t,x)]^{2}}{\frac{\partial^{2} v}{\partial x^{2}}(t,x)}.$$

Similarly
$${\lim_{\epsilon\downarrow 0}}\frac{\mathbb{E}[v(t+\epsilon,{X}^{t,x}(t+\epsilon)-v(t,x)]}{\epsilon}=$$
$${\lim_{\epsilon\downarrow 0}}\frac{\mathbb{E}\int_{t}^{t+\epsilon}d v(u, {X}^{t,x}(u))}{\epsilon}=
\frac{\partial v}{\partial t}(t,x)+\left[(r+\mu\zeta-c)x\frac{\partial v}{\partial x}(t,x)+\frac{1}{2}\sigma^{2}\zeta^{2}x^{2}\frac{\partial^{2} v}{\partial x^{2}}(t,x)\right].$$

$3.$ In the light of inequalities \eqref{089}, \eqref{189} and Dominated Convergence Theorem it follows that

$${\lim_{\epsilon\downarrow 0}}\frac{\mathbb{E}\left[\int_{t+\epsilon}^{T}[h(s-t)-h(s-t-\epsilon)][U(F_2(s,\bar{X}^{t,x}(s)))-U(F_2(s,{X}^{t,x}(s)))]\,ds\right]}{\epsilon}=0.$$

Therefore

$${\lim_{\epsilon\downarrow 0}}\frac{J(t,x,F_{1},F_{2})-J(t,x,\zeta_{\epsilon},c_{\epsilon})}{\epsilon}=$$
$$\left[\frac{\partial v}{\partial t}(t,x)+\left(rx-I\left(\frac{\partial v}{\partial x}(t,x)\right)\right)\frac{\partial v}{\partial x}(t,x)-
\frac{\mu^{2}}{2\sigma^{2}}\frac{{[\frac{\partial v}{\partial x}}(t,x)]^{2}}{\frac{\partial^{2} v}{\partial x^{2}}(t,x)}+U(F_2(t,x)) \right]- $$
$$\left[\frac{\partial v}{\partial t}(t,x)+\left[(r+\mu\zeta-c)x\frac{\partial v}{\partial x}(t,x)+\frac{1}{2}\sigma^{2}\zeta^{2}x^{2}\frac{\partial^{2} v}{\partial x^{2}}(t,x)\right]+U(cx)\right]$$ $$\geq0,$$

where the last inequality comes from \eqref{io}

\begin{flushright}
$\square$
\end{flushright}

{\bf G Proof of Theorem \ref{ain2}}

 The proof of this Theorem will be split into several Lemmas.

\begin{lemma}\label{in2}
Assume there exists a solution of the integral equation \eqref{IE0}. Then there exists a  positive constant $A$ (that depends on $T$
 and $h(\cdot)$) such that  

\begin{equation}\label{&12}
\inf_{t\in[0,T]}\lambda(t)> e^{-A T},\qquad \sup_{t\in[0,T]}\lambda(t)< \left[\frac{1-p}{A} +\left(\frac{1-p}{A}+1\right)e^{\left(\frac{A T}{1-p} \right)} \right]^{1-p}.
 \end{equation}

\end{lemma}

{\bf  Proof}

The equation \eqref{IE0} can be written in differential form as

\begin{eqnarray}\label{&123}
\lambda'(t)&=&-\left[\frac{h'(T-t)}{h(T-t)}+K\right]\lambda(t)+ (p-1)[\lambda(t)]^{\frac{p}{p-1}}\\\notag&+&\int_{t}^{T} h(T-t)\frac{\partial}{\partial t}\left[\frac{h(s-t)}{h(T-t)}\right][\lambda(s)]^{\frac{p}{p-1}}e^{-\left(\int_{t}^{s}p[\lambda(u)]^{\frac{1}{p-1}}\,du\right)}\,ds.
 \end{eqnarray}
From the equation \eqref{IE0} we infer that

$$\int_{t}^{T} h(s-t)e^{K(s-t)} [\lambda(s)]^{\frac{p}{p-1}}e^{-\left(\int_{t}^{s}p[\lambda(u)]^{\frac{1}{p-1}}\,du\right)}\,ds\leq \lambda(t).$$

This together with the boundedness (from below and above) of the function $h,$ yields a positive constant $A$ (that depends on $T$
 and $h(\cdot)$) such that

\begin{eqnarray}\label{&124}
\lambda'(t)&\leq& A\lambda(t)+ (p-1)[\lambda(t)]^{\frac{p}{p-1}},
 \end{eqnarray}
and

\begin{eqnarray}\label{&125}
\lambda'(t)&\geq&  -A\lambda(t)+ (p-1)[\lambda(t)]^{\frac{p}{p-1}}.
 \end{eqnarray}

It follows from \eqref{&124} that
$$\vartheta'(t)-\left(\frac{A}{1-p} \right)\vartheta(t)\leq -1,$$
where
$$\vartheta(t)\triangleq\left(\frac{1}{[\lambda(t)]^{\frac{1}{p-1}}} \right). $$ Thus
$$\left(e^{\left(\frac{At}{p-1}\right)}\vartheta(t) \right)'\leq-e^{\left(\frac{At}{p-1}\right)}<0.$$ Integrating this from $t$ to $T$
we obtain the lower estimate on $\lambda(t).$ Following  \eqref{&124} and arguing similarly we can get the upper  estimate on $\lambda(t).$

\begin{flushright}
$\square$
\end{flushright}

The next Lemma establishes uniqueness of  \eqref{IE0}. 
\begin{lemma}\label{in3}
There is a unique solution to \eqref{IE0}.
 
\end{lemma}

{\bf  Proof}

We argue by contraposition. Let $\lambda_1$ and $\lambda_2$ two solutions of \eqref{IE0}. Then in the light of the estimate of
Lemma \ref{in2} it follows that
$$\bigg|h(T-t)e^{K(T-t)}e^{-\left(\int_{t}^{T}p[\lambda_{1}(u)]^{\frac{1}{p-1}}\,du\right)}-h(T-t)e^{K(T-t)}e^{-\left(\int_{t}^{T}p[\lambda_{2}(u)]^{\frac{1}{p-1}}\,du\right)} \bigg|\leq K_1 \int_{t}^{T}|\lambda_{1}(u)-\lambda_{2}(u)|\,du  $$

and

$$|[\lambda_{1}(t)]^{\frac{p}{p-1}}-[\lambda_{2}(t)]^{\frac{p}{p-1}}|\leq K_2 |\lambda_{1}(t)-\lambda_{2}(t)|  ,$$

for some positive constants $K_1$ and $K_2.$ Therefore there exists a positive constant $K_3$ such that

$$|\lambda_{1}(t)-\lambda_{2}(t)|\leq K_3 \int_{t}^{T}|\lambda_{1}(u)-\lambda_{2}(u)|\,du.  $$

At this point one can argue as in the Gronwal's Lemma proof to get that $|\lambda_{1}(t)-\lambda_{2}(t)|=0.$

\begin{flushright}
$\square$
\end{flushright}

Next we show existence of a solution of \eqref{IE0} if the discount function $h(t)$ is a combination of exponentials.

\begin{lemma}\label{in3}
Let the discount function $h(t)$ be of the form $h(t)=\sum_{n=1}^{N}\beta_{n}e^{-\rho_{n} t}$ for some positive constant $\beta_{n}$
such that $\sum_{n=1}^{N}\beta_{n}=1.$ Then the equation \eqref{IE0} has a unique solution.  
\end{lemma}

{\bf  Proof}

The uniqueness is a consequence of Lemma \ref{in3}. As for existence let us notice that the integral equation \eqref{IE0} can be reduced in this case
to a system of ODE. Indeed let us denote $h_n(t)=e^{-\rho_{n} t}.$ Then
\begin{equation}\label{IE0n}
\begin{cases}
\lambda_{n}(t)=\int_{t}^{T} h_{n}(s-t)e^{K(s-t)} [\lambda(s)]^{\frac{p}{p-1}}e^{-\left(\int_{t}^{s}p[\lambda(u)]^{\frac{1}{p-1}}\,du\right)}\,ds+
h_{n}(T-t)e^{K(T-t)}e^{-\left(\int_{t}^{T}p[\lambda(u)]^{\frac{1}{p-1}}\,du\right)}\\\lambda_{n}(T)=1\end{cases}
\end{equation}

solves the following ODE system

\begin{equation}\label{IE0nn}
\begin{cases}
\lambda'_{n}(t)=(\rho_{n}-K)\lambda_{n}(t)+(p-1)[\lambda(t)]^{\frac{p}{p-1}}\\\lambda_{n}(T)=1\\\lambda(t)=\sum_{n=1}^{N}\beta_{n}\lambda_{n}(t)\end{cases}
\end{equation}
Existence of a solution for this system is granted locally by The Schauder Fixed Point Theorem. However the estimates given by Lemma
\ref{in2} grant a solution globally.

\begin{flushright}
$\square$
\end{flushright}

{\bf G Proof of Theorem \ref{ain2}} 

Let $\mathcal{C}=\{h:[0,T]\rightarrow \mathbb{R} |\,\,h(t)=\sum_{n=1}^{N}\beta_{n}e^{-\rho_{n} t}\}$
then $\mathcal{C}$ is dense in $C([0,T],\mathbb{R}).$ Thus the function $h'(t)$ can be approximated by functions in $\mathcal{C}$.
Moreover $h(t)$ an be approximated (in the sup norm) by $h^{n}(t)\in \mathcal{C}$ such that $h'(t)$ is approximated (in the sup norm) by $[h^{n}]'(t).$ From $\inf_{t\in[0,T]} h(t)>0$ it follows that for big enough $n,$ $\inf_{t\in[0,T]} h_{n}(t)>0.$ Since
$$\frac{h^{n}(T-t)}{h^{n}(s-t)}\frac{\partial}{\partial t}\left[\frac{h^{n}(s-t)}{h^{n}(T-t)}\right] \longrightarrow \frac{h(T-t)}{h(s-t)}\frac{\partial}{\partial t}\left[\frac{h(s-t)}{h(T-t)}\right] $$
 as ${n\rightarrow\infty},$ one can find for big enough $n$ a constant $A$ such that the estimates of Lemma \ref{in2} hold with the same $A$
 (independent of $n$) if one considers the integral equation \eqref{IE0} with $h(t)=h^{n}(t)\in \mathcal{C}.$ The solution $\lambda^{n}(t)$ is uniformly bounded in $n$. Moreover from \eqref{&123} we see it has uniformly bounded derivative so it is equicontinuous. Thus, Arzela-Ascoli compactness criterion grants convergence of  $\lambda^{n}(t)$  as ${n\rightarrow\infty}$ to some function $\lambda(t).$ Passing at limit in \eqref{IE0} with $h(t)=h^{n}(t)\in \mathcal{C}$ using Dominated Convergence Theorem (with the estimates of Lemma \ref{in2} holding with the same $A$ for all $n$) we get that $\lambda(t)$ solves \eqref{IE0}.

\begin{flushright}
$\square$
\end{flushright}

\end{document}